\documentclass{article}
\usepackage[preprint]{neurips_2024}
\usepackage{longtable}
\usepackage[T1]{fontenc}
\usepackage[notext]{stix2}
\usepackage[table,svgnames]{xcolor}
\usepackage{microtype}
\usepackage{hyperref}
\usepackage{url}
\usepackage{booktabs}
\usepackage{graphicx} 
\usepackage{amsmath}
\usepackage{amsthm}
\usepackage{algorithm}
\usepackage{algpseudocode}
\usepackage{bm}
\usepackage{cleveref}
\usepackage{tikz}
\usepackage{pgfplotstable}
\usepackage{subcaption}
\usepackage{epigraph}
\usepackage{enumitem}
\usepackage{multirow}
\usepackage[font=small]{caption}

\usepackage{makecell}

\usetikzlibrary{positioning,fit,calc}

\usepackage{xcolor}
\hypersetup{%
    colorlinks,
    linkcolor={red!50!black},
    citecolor={blue!50!black},
    urlcolor={blue!80!black}
}
\pgfplotsset{compat=1.18}

\setlist[description]{%
  topsep=1ex,     
  itemsep=1ex,     
  font={\normalfont\itshape}, 
  leftmargin=0ex,
}

\title{{\it Sense and Sensitivity:} Evaluating the simulation of\\social dynamics via Large Language Models}
\author{
  Da Ju \\
  Meta FAIR \\
  \texttt{daju@meta.com} \\
  \And
  Adina Williams \\
  Meta FAIR \\
  \texttt{adinawilliams@meta.com} \\
  \And
  Brian Karrer \\
  Meta FAIR \\
    \texttt{briankarrer@meta.com} \\
    \And
      Maximilian Nickel \\
  Meta FAIR \\
    \texttt{maxn@meta.com} 
}

\renewcommand{\vec}[1]{\bm{#1}}
\newcommand{\vx}{\vec{x}}
\newcommand{\Set}[1]{\mathcal{#1}}
\newcommand{\E}{\mathbb{E}}
\newcommand{\R}{\mathbb{R}}
\newcommand{\T}{\mathbb{T}}
\newcommand{\draw}{\sim}
\newcommand{\loss}{\mathcal{L}}
\newcommand{\dset}{\Set{D}}
\newcommand{\Eerr}{\E_{X_T \draw p_{X|T=t}}\left[\loss(f(X_T), h(X_T))\right]}
\newcommand{\Nset}{\Set{N}}
\newcommand{\llama}{\textsc{LLaMA}}

\newcommand{\goodcell}{\cellcolor{green!25}}
\newcommand{\badcell}{\cellcolor{red!25}}

\algnewcommand\algorithmicforeach{\textbf{for each}}
\algdef{S}[FOR]{ForEach}[1]{\algorithmicforeach\ #1\ \algorithmicdo}

\pgfdeclarelayer{bg}    
\pgfsetlayers{bg,main}

\theoremstyle{definition}
\newtheorem{definition}{Definition}

\begin{document}

\maketitle

\begin{abstract}
    Large language models have increasingly been proposed as a powerful replacement for classical agent-based models (ABMs) to simulate social dynamics. By using LLMs as a proxy for human behavior, the hope of this new approach is to be able to simulate significantly more complex dynamics than with classical ABMs and gain new insights in fields such as social science, political science, and economics. However, due to the black box nature of LLMs, it is unclear whether LLM agents actually execute the intended semantics that are encoded in their natural language instructions and, if the resulting dynamics of interactions are meaningful. To study this question, we propose a new evaluation framework that grounds LLM simulations within the dynamics of established reference models of social science. By treating LLMs as a black-box function, we evaluate their input-output behavior relative to this reference model, which allows us to evaluate detailed aspects of their behavior. Our results show that, while it is possible to engineer prompts that approximate the intended dynamics, the quality of these simulations is highly sensitive to the particular choice of prompts. Importantly, simulations are even sensitive to arbitrary variations such as minor wording changes and whitespace. This puts into question the usefulness of current versions of LLMs for meaningful simulations, as without a reference model, it is impossible to determine a priori what impact seemingly meaningless changes in prompt will have on the simulation.
\end{abstract}


\section{Introduction}\label{sec:intro}

Generative social science aims to explain social phenomena through agent-based models (ABMs), i.e., through models of a large number of interacting agents. By being able to \emph{generate} macroscopic phenomena of interest through simple, interpretable micro-instructions to agents, these models aim to provide insights into the mechanisms that are sufficient to explain social phenomena~\citep{Epstein1997-jo}. 
Early ABMs have indeed seen substantial success for this task, ranging from \citeauthor{Schelling1971-qd}'s \citeyearpar{Schelling1971-qd} seminal work on segregation to \citeauthor{Brian-Arthur2018-ck}'s \citeyearpar{Brian-Arthur2018-ck} modeling of the stock market. Agent-based models have since found widespread use in social science~\citep{Gilbert2000-iq}, economics~\citep{Tesfatsion2023-af}, finance~\citep{Chan1999-ou}, epidemiology~\citep{Auchincloss2008-ok}, public health~\citep{Tracy2018-hb}, AI~\citep{Zhang2021-ns}, and many other fields. 
However, despite these successes, classical ABMs also face a variety of substantial challenges~\citep{An2021-ck}. For the purposes of this paper, the most relevant of these challenges include:
\begingroup
\advance\leftmargini -2.5em
\begin{quote}
\begin{description}
    \item[Highly idealized assumptions and models:] Since classical ABMs need to abstract behavior into highly idealized mathematical models, a core challenge is to create sufficiently rich models that can be considered realistic proxies for actual human behavior. For instance, models for belief formation and opinion dynamics represent opinions of agents towards a topic as a single scalar which evolves through interactions with other agents. While this leads to interpretable models, they also fail to consistently reproduce dynamics that are observed in the real-world.
    \item[Susceptibility for misspecification:] Since classical ABMs are not grounded in human behavior or data, it is entirely a modeler's responsibility to accurately and comprehensively capture the relevant behavior for a task in their models. Since emergent macroscopic behavior, which is the core focus of ABMs, can be very sensitive to the right kind of micro-instructions, the danger of wrong inferences due to model misspecification is substantial in classical ABMs.
    \item[Difficulty of empirical validation:] Again, due to their abstraction level, ABMs often do not have clear correspondences to real-world phenomena. It can therefore be quite difficult to measure and understand whether a model captures real phenomena or is misspecified.
\end{description}
\end{quote}
\endgroup
The slow progress and difficulty of addressing these issues related to methodology and validation have raised serious questions about the utility of classical ABMs. Subsequently this led to a decreasing interest in these models in various fields \citep{An2021-ck,Roughgarden2012-ae, Couclelis2001-bw}.

As LLMs have increasingly shown that they can act as proxies for human behavior, at least in some aspects, there has been substantial interest in utilizing them for agent-based models to overcome these challenges~\citep{Park2023-ur,Gao2023-fm,Tornberg2023-bt,Kaiya2023-wt}. 
Various aspects of LLMs are highly promising for this purpose: (i) LLMs allow to simulate rich, complex behaviors which are grounded in the training corpora of these models.~(ii) LLMs can show new emergent behavior --- due to being sufficiently rich --- that has not been explicitly encoded in the model (which logical ABMs are not able to).~(iii)	Specifying micro-instructions in natural language allows for richness and complexity, while remaining interpretable such that human agents can even be part of a simulation.
Successfully applying LLMs to ABMs could therefore be a transformative technology for social science by creating high-fidelity simulations of social systems. This is especially the case for social systems where LLMs can be expected to have substantial training data such as social media and internet platforms.
Consequently, research on LLM-ABMs has increasingly considered simulating opinion dynamics~\citep{Chuang2023-uq} and effects of ranking algorithms in social media~\citep{Tornberg2023-bt}.
However, given the aims of ABMs in science --- to produce reliable insights and support theory-building --- are the insights that we can get from LLM-ABMs actually reliable and robust? Due to the black-box nature of LLMs it is entirely unclear how \emph{exactly} an LLM agent interprets its instructions and acts upon it. This is in complete contrast to classical ABMs in which the micro-behavior is directly encoded in the model.
Moreover, it is well-known that current LLMs can be highly sensitive to the choice of prompts~\citep{Wallace2019-kh, reynolds-etal-2021-prompt, min-etal-2022-rethinking, webson-pavlick-2022-prompt, goodarzi-etal-2023-robustness, lyu-etal-2024-keeping, shivagunde-etal-2024-deconstructing-context}, to the extent that \emph{prompt engineer} is now a job title in the field of AI. Under these circumstances, how can we know whether a seemingly minor change in a prompt will not lead to a substantial change in the simulation outcome? Even if we have identified a set of prompts that simulates an intended system well, can we really consider these prompts as an interpretable model if minor changes lead to significantly different outcomes?
If LLM-ABMs shall ever be useful for theory-building and analysis in \emph{science}, it is therefore of highest importance to establish validation frameworks that allow us to examine two \emph{necessary} conditions for an ABM to be useful for scientific analyses, i.e., (a) their consistency with the expected dynamics and (b) their reliability. 
For this purpose, we introduce a new evaluation framework that grounds a LLM-ABM in the dynamics of a reference ABM to analyze its behavior. Specifically, we will use an approach akin to system identification in control theory, where we treat the LLM as a black-box function mapping an input state to an output state. We then evaluate the simulation error, i.e, the input-output behavior of the LLM relative to a reference model of social dynamics what allows us to evaluate detailed aspects of the behavior of the LLM-ABM. While this is a general framework, we will focus in this work on two specific aspects in this evaluation: (a) whether LLM-ABMs can reproduce the dynamics of classical ABMs and (b) how sensitive LLM-ABMs are with respect to minor variations in prompts. 


\section{Agent-Based Models of Social Dynamics}\label{sec:abm}
\begin{figure}
    \begin{subfigure}[t]{.5\columnwidth}
        \centering
        \includegraphics[width=.85\columnwidth]{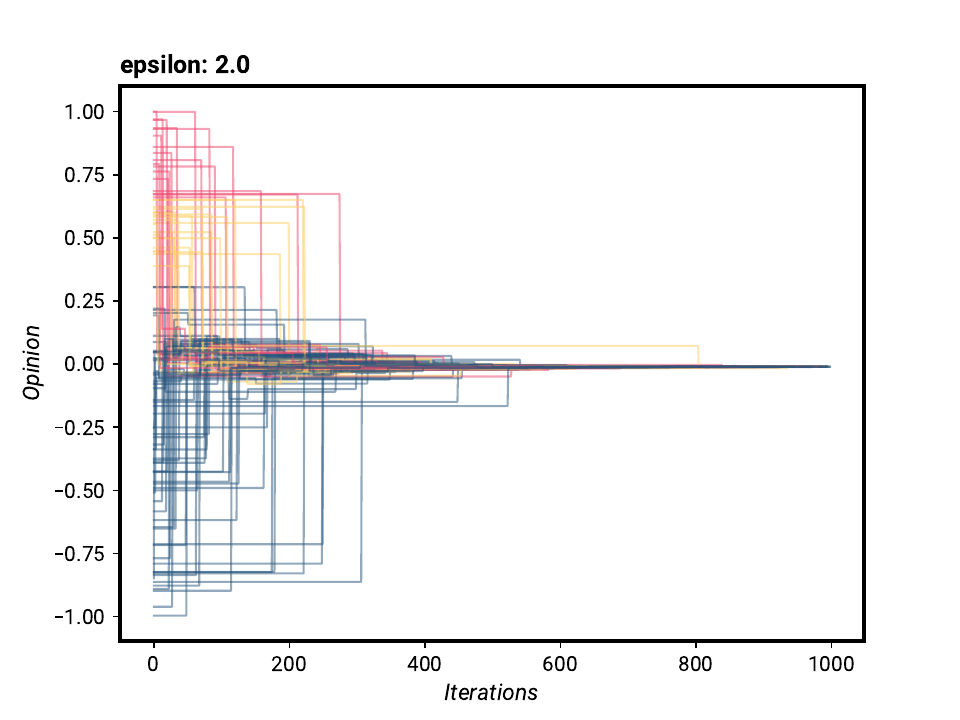}
        \caption{DeGroot}\label{fig:degroot}
    \end{subfigure}%
    \begin{subfigure}[t]{.5\columnwidth}
        \centering
        \includegraphics[width=.85\columnwidth]{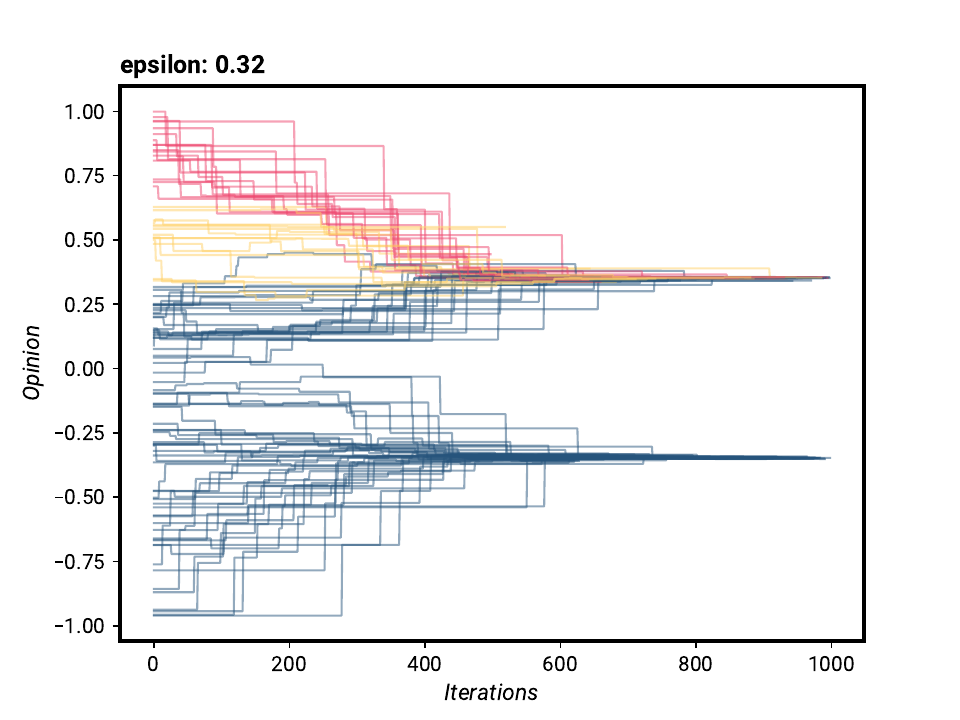}
        \caption{Hegselmann-Krause}\label{fig:hk}
    \end{subfigure}
    \caption{\textit{Opinion Dynamics Models}. Scalar opinion \(x_i(t)\) per agent \(i\) and time \(t\) for the (\subref{fig:degroot}) DeGroot and (\cref{fig:hk}) Hegselmann-Krause models, simulated using \cref{eq:hk} with \(\epsilon=2\) and \(\epsilon=0.3\) respectively.  While the DeGroot model leads to consensus and the main quantity of interest is the speed of convergence, the Hegselmann-Krause model can show a wider range of dynamics, including polarization as in \cref{fig:hk}.}\label{fig:opdyn}
\end{figure}

Agent-based models serve a similar purpose in the social sciences as computational simulations do in the natural sciences such as physics, chemistry, and cosmology. In particular, ABMs have played important roles in theory-building, constructing models, and simulation for understanding complex dynamics in a wide range of social processes, ranging from segregation, to financial markets, to prehistoric migrations. Many such agent-based models can be described in the following form: Let \(N\) denote the number of agents and let \(\vx(t) \in \R^N\) denote the state of each agent at time \(t\). Furthermore, let \(f : \smash{\R^N} \to \smash{\R^N}\) denote the dynamics of the system, i.e., the function that describes how the state of the system evolves over time such that 
\begin{equation}
    \vx(t+1) = f(\vx(t)) .
\end{equation}
Importantly, \(f\) is a function of the \emph{entire} state vector and can thus model interactions between agents. 
To endow \(f\) with meaning, we will focus in the following mainly on a key application area of ABMs that coincides well with the strengths of LLMs, i.e., the simulation of opinion dynamics.
Classical ABMs for simulating opinion dynamics are structured as follows: First, \(\vx(t) \in \R^N\) denotes the vector of opinions of all agents towards a specific topic where the exact encoding of opinions as a scalar can be application- and model-specific. Typical encoding schemes include \(x_i \in [-1,1]\) or \(x_i \in [0, 1]\) for indicating opinions that strongly support (\(1\)) to strongly oppose (\(-1\) or \(0\)) a particular topic.
Within this framework, one of the earliest models for simulating opinion dynamics via agent-based models has been proposed by \citet{DeGroot1974-zh}, which defines \(f\) as follows:

\begin{definition}[DeGroot Model]\label{def:degroot}
Let \(G = (V,E)\) denote a social network such that $V$ corresponds to a population of agents and \(E\) denotes the existing relationships (edges) between them. Furthermore, let \(\Nset(i) = \{j : i \sim j \in E\}\) denote the set of all neighbors of agent \(i \in V\) in \(G\). Each agent \(i\) updates then their opinion at time \(t\) by averaging the opinions of its neighbors, i.e.,
\setlength{\belowdisplayskip}{3pt}%
\begin{equation}
    f{(x(t))}_i = \textstyle\frac{1}{|\Nset(i)|} \sum_{j \in \Nset(i)} x_j(t)
\end{equation}
\end{definition}
While the DeGroot model has been an influential model for consensus formation and social learning, its use for simulating general opinion dynamics is limited. This is because, under a wide range of initial conditions, this model can be proven to always converge to a consensus, and in particular to the average of the initial conditions. To understand opinion dynamics in the context of polarization and radicalization, more flexible models are therefore needed. 
One of the most impactful extensions of DeGroot's models in this regard has been proposed by \citet{Hegselmann2002-qh}. This model shows that a wide range of dynamics including polarization can arise through a surprisingly simple addition to DeGroot's model, i.e., \emph{bounded confidence}, which says that agents will not consider any opinions in their updates that are too far away from their own. This model is defined as follows:
\begin{definition}[Hegselmann-Krause Model]\label{def:hk}
Let \(G = (V, E)\) denote a social network and let 
\begin{equation}
    \Nset(i,t) = \{ j : i \sim j \land |x_i(t) - x_j(t)| \leq \epsilon \}
\label{eq:epsilon}
\end{equation}
denote the set of neighbors of agent \(i\) within confidence bound \(\epsilon\). Each agent \(i\) updates then their opinion at time \(t\) by averaging the opinions of neighbors within their confidence bound, i.e.,
\setlength{\belowdisplayskip}{3pt}%
\begin{equation}
    f{(x(t))}_i = \textstyle\frac{1}{|\Nset(i,t)|}\sum_{j \in \Nset(i,t)} x_j(t)
\label{eq:hk}
\end{equation}
\end{definition}
See \cref{fig:opdyn} for an illustration of the main differences in opinion dynamics between a DeGroot and a Hegselmann-Krause model, i.e., between consensus and polarization. 
%
Next, we will introduce our proposed framework to measure the consistency of an LLM-based ABM with such reference models.

\begin{figure}[t]
    \begin{subfigure}{.33\linewidth}
        \centering
        \raisebox{1em}{%
            \includegraphics[width=.95\linewidth]{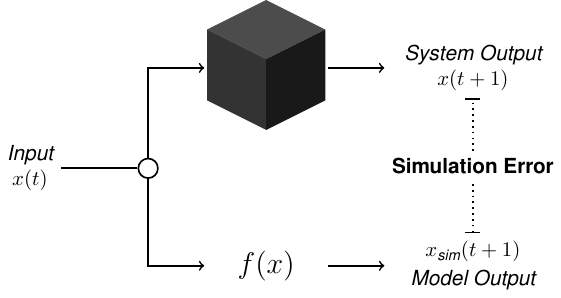}
        }
        \caption{Simulation Error}\label{fig:simerr}
    \end{subfigure}%
    \begin{subfigure}{.67\linewidth}    
        \centering
        \resizebox{.95\linewidth}{!}{%
        \begin{tikzpicture}[thick,font=\small\sffamily]
        \node (x) {$x(t)$};
        \node[right=6 of x.center] (fx) {$x(t+1)$};
        \node[above=2 of x] (ex) {$y(t)$};
        \node[right=6 of ex.center] (hx) {$y(t+1)$};
        \node[right=1 of fx] (ref) {\sffamily\textbf{Reference model}};
        \node[right=1 of hx] (llm) {\sffamily\textbf{LLM model\hspace*{1.95em}}};
        \draw[->] (x) -- (ex) node[midway,fill=white,align=center] (enc) {encode\\$e(x)$};
        \draw[->] (hx) -- (fx) node[midway,fill=white,align=center] {decode\\$d(x)$};
        \draw[->] (x) -- (fx) node[midway,fill=yellow!30,align=center,yshift=-0.6em] (rdl) {Reference dynamics\\$f(x)$};
        \draw[->] (ex) -- (hx) node[midway,fill=white,align=center,yshift=-0.6em] (sdl) {LLM dynamics\\$g(x)$};

        \begin{pgfonlayer}{bg}
            \node[fit=(enc)(ex)(hx)(llm)(sdl),inner sep=1ex,draw=black,rounded corners, ultra thick, dotted] {};
            \node[fit={([xshift={-.9em}]x.west)(fx)(ref)(rdl)},inner sep=1ex,fill=yellow!30,rounded corners] {};
        \end{pgfonlayer}
        \end{tikzpicture}
        }
        \caption{Model consistency}\label{fig:scheme}
    \end{subfigure}
    \caption{\textit{Evaluation Framework}. (\subref{fig:simerr}) Simulation error in black-box system identification (\subref{fig:scheme}) Schematic of the evaluation framework, showing encoding, decoding and dynamics operations.}\label{fig:eval}
\end{figure}
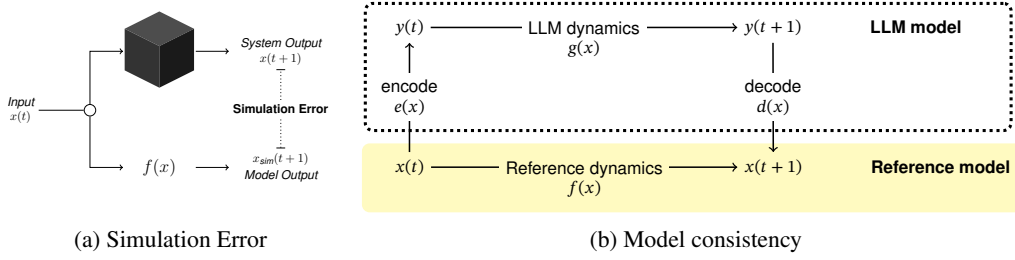
\section{Method}
\label{sec:method}
Given a reference model as in \cref{sec:abm}, we can now define a framework to measure the two \emph{necessary conditions} that we are concerned with: (a) how consistent an LLM-ABM is with the dynamics of the reference model and (b) how robust the LLM-ABM is to minor variations in prompting.
To approach these questions, we will adopt an approach similar to system identification in control theory, and evaluate the \emph{simulation error} of the LLM-ABM relative to the reference model~\citep{Ljung1998-qx}. 
%
In particular, let \(\Set{X}\) denote the domain of the \emph{population state}, e.g., \(\R^N\) or \([-1,1]^N\) for population opinions. Furthermore, let \(p_{X|T}\) be the conditional distribution of the population state at time \(T\). Such stochasticity in the dynamics can be introduced via random initial conditions \(x(0)\) or through the transition function \(f\), as for instance in \citep{Schelling1971-qd}. Next, let \(\theta\) be a set of prompts, let \(\Set{M}\) be a pre-trained LLM such as \llama~\citep{Touvron2023-gx}, and let \(h_{\Set{M},\theta} : \Set{X} \to \Set{X}\) be a LLM-ABM parametrized by \(\theta\) that maps a population state \(\vx(t)\) to its successor state. Since we will only consider models over a single pre-trained LLM, we will drop the subscript \(\Set{M}\) in the following for conciseness.
Next, to compare the dynamics of LLM-ABMs relative to a reference ABM, we need to address the obvious domain mismatch, i.e., the fact  that LLMs don't operate natively in \(\Set{X}\), but rather in the space of natural language (in the following denoted by \(\T\)). Due to this issue, the two dynamics are not directly comparable. To overcome this, we propose the following encoding-decoding scheme (see also \cref{fig:scheme}):
\begin{align}\label{eq:encdec}
    \text{Encoder~:} \quad & e_{\theta_\text{enc}} : \R \to \T & & &
    \text{LLM dynamics~:} \quad & g_{\theta_\text{dyn}} : \T \to \T \\
    \text{Decoder~:} \quad & d_{\theta_\text{dec}} : \T \to \R \notag & & &
    \text{LLM model~:} \quad & h_\theta(x) = d_{\theta_\text{dec}}\ \circ\ g_{\theta_\text{sim}}\ \circ\ e_{\theta_\text{enc}}(x) \notag
\end{align}
In \cref{eq:encdec}, we first encode a scalar state value into a natural language representation, then apply the LLM dynamics to the natural language representation and finally decode this back into a scalar state. All components \(e\), \(g\), \(d\) are parametrized by \(\Set{M}\) and instruction prompts that specify the encoding-decoding operations and the dynamics itself. For \(h\) it holds simply that \(\theta = \theta_\text{enc} \cup \theta_\text{dec} \cup \theta_\text{dyn}\). 

Given this encoding-decoding scheme, we can now introduce the general notion of 
consistency of a candidate LLM-ABM with a pre-defined reference model:
\begin{definition}[Model inconsistency]\label{def:mcons}
 Let \(f : \Set{X} \to \Set{X}\) denote a reference model and \(h_\theta : \Set{X} \to \Set{X}\) an candidate model. Let \(\loss : \Set{X} \times \Set{X} \to \R_+\) denote a loss function. The \emph{inconsistency of \(h_\theta\) with \(f\)} is then defined as the \emph{expected simulation error} over \(p(x|t)\) averaged over all time step \(t\), i.e., 
\begin{equation}\label{eq:mcons}
    C(f, h_\theta) = \textstyle \frac{1}{T}\sum_{t=1}^T\Eerr = \frac{1}{T} \sum_{t=1}^T \int_\Set{X} \loss(f(x), h_\theta(x))\ p(x \mid t)\ dx 
\end{equation}
\end{definition}
Since we have, by definition, access to the reference model, we can use Monte-Carlo estimation to approximate \cref{eq:mcons}. In particular, let \(\dset = \smash{{\{(\vx(t_k))\}}\vphantom{m}^K_{k=1}}\) be a dataset that is sampled from \(p(x|t)\) by simulating the reference model forward in time. Then, \cref{eq:mcons} can be approximated via
\begin{equation}
    \Eerr \approx \textstyle \frac{1}{|\dset|}\sum_{x \in \dset}\loss(f(x), h(x))
    \label{eq:monte}
\end{equation}

Given \cref{def:mcons} and \cref{eq:monte}, we can now evaluate the consistency of a LLM-ABM with a reference model.
However, an important aspect to note is that \cref{eq:encdec} introduces an additional source of error that might not be due to the LLM dynamics itself, but instead to the encoding-decoding operations. For this reason, it is important to also measure the consistency of these operations when combining \cref{eq:encdec} with \cref{def:mcons}, i.e., that an encoded scalar is decoded back to a similar value. We define this form of consistency as follows:
\begin{definition}[Encoding-Decoding consistency]\label{def:edcons}
Let \(e : \R \to \T\) and \(d : \T \to \R\) be encoding and decoding operations, respectively and let \(\loss : \R \times \R \to \R_+\) denote a loss function. Then, the consistency of these operations is defined as the expected loss over \(p_X\) between a scalar opinion value and the chained encoding-decoding operation, i.e.,
\begin{equation}
    \smash{\E_{X \draw p_X}}[\loss(X, d\ \circ\ e(X)]
\end{equation}
\end{definition}
%

Given a computable consistency measure as in \cref{def:mcons}, we can now also proceed to define the sensitivity of a LLM-ABM to prompt variations. In slight abuse of notation, let \(\theta + \epsilon\) denote the (slight) variation of a prompt, e.g., semantically equivalent phrasings or adding of whitespace. 
\begin{definition}[Model sensitivity]\label{def:sens}
    Let \(f\) and \(h_\theta\) be reference and candidate model, respectively. Further, let \(h_{\theta + \epsilon}\) denote the candidate model with a slight variation in prompts. The sensitivity of \(h_\theta\) to prompt variations is then defined as the absolute difference in consistency between the models,    
    \begin{equation}
        S(f, h_\theta, \epsilon) = |C(f, h_\theta) - C(f, h_{\theta + \epsilon})|
    \end{equation}
\end{definition}
Given \cref{def:mcons,def:sens}, we now evaluate the consistency of a LLM-ABM and its sensitivity to prompt variations. Without being able to achieve consistency, it is clear that the LLM-ABM is not able to simulate the reference model sufficiently and any observed prompt variations are not informative in this case. With regard to sensitivity, the interpretation is more subtle. If the sensitivity is high, it means that the LLM-ABM is highly sensitive to the choice of prompts. Under this circumstances, it is hard to argue that the LLM-ABM fulfills its intended use of providing interpretable insights into the dynamics since an almost equal explanation leads to a substantially different outcome.

\section{Simulating LLM-Agent Opinion Dynamics at Scale}\label{sec:sim}
To give meaning to \cref{sec:method}, we need to instantiate the social network \(G\), the encoding-decoding pair \(e, d\), as well as the LLM dynamics \(g\) with concrete objects. In the following, we will discuss options for these choices in the context of simulating opinion dynamics as in \cref{sec:abm}. Furthermore we will discuss how such simulations can be achieved at scale, i.e., for thousands of agents.

\vspace{-0.75\baselineskip}
\paragraph{Social network} The social network \(G\) can be instantiated either according to real-world social networks ranging from social media~\citep{Adamic2005-zm,Backstrom2006-cr} over academia~\citep{Leskovec2005-qg} to congress~\citep{Fink2023-yc}, or, alternatively, via random social network generators such as stochastic block models~\citep{Holland1983-ew,Karrer2011-ri} and the Barabasi-Albert model \citep{Barabasi1999-ua}. It is well-known that different network structures (e.g., scale-free versus small-world as in~\citealt{Watts1998-db}) can lead to different opinion dynamics. However, since we will not focus on the effect of network structure in this work, we will instantiate \(G\) simply via a standard Erd\H{o}s-R{\'e}nyi model \(G(n,p)\) \citep{Erdos1960-mv} where \(n\) denotes the number of nodes and \(p\) denotes the probability of an edge between in the graph. Hence, edges in this model are created randomly according to
\(
    \forall i,j \in V: \Pr(i \sim j) = p .
\)
\vspace{-0.75\baselineskip}
\paragraph{Agent Personas} To fulfill the promise of LLMs to create more realistic social dynamics, each agent needs also to be instantiated with realistic personas. However, achieving this at scale is non-trivial. To access a large and diverse range of personas, we utilize the crowd-sourced \textsc{PersonaChat} data set \citep{Zhang2018-hb}. To infuse agents with their respective personas, we prefix each prompt with the persona information for this agent. See \cref{tab:prompts} in appendix \ref{app:prompt_example} for examples. 
\vspace{-0.75\baselineskip}
\paragraph{Agent micro-instructions} Given a social network and agent personas, we can then define the ABM via instruction prompts. In particular, the \emph{encoding prompt} \(\theta_\text{enc}\) takes as input a scalar opinion about a topic as well as the agent's persona and outputs a text that reflects the agent's opinion. The \emph{dynamics prompt} \(\theta_\text{dyn}\) takes as input the agent's persona and a timeline of posts from the agent's network and outputs the agent's updated opinion. The \emph{decoding prompt} \(\theta_\text{dec}\) takes as input the agent's updated opinion and outputs a scalar opinion. See \cref{tab:prompts} in appendix \ref{app:prompt_example} for examples of these prompts.
\begin{figure}[t]
    \begin{subfigure}{.3\linewidth}
        \centering
        \includegraphics[width=.9\linewidth]{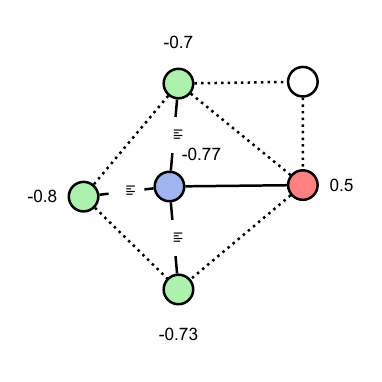}
        \caption{Ego Network}\label{fig:ego}
    \end{subfigure}%
    \begin{subfigure}{.7\linewidth}
        \centering
        \raisebox{1em}{\includegraphics[width=.9\columnwidth]{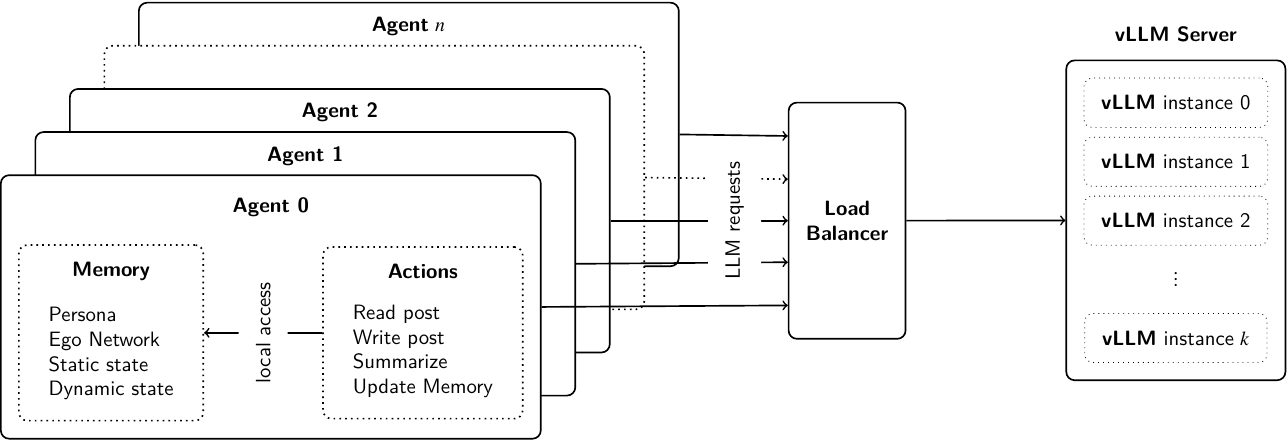}}
        \caption{System architecture}\label{fig:arch}
    \end{subfigure}
    \caption{(\subref{fig:ego}) \textit{Ego network and opinion aggregation}. The ego network of node Blue is indicated by solid edges. Nodes in Blue's ego network that are within an \(\epsilon\) distance of Blue's opinion are indicated in green. Edges which transmit opinions are indicated by a text icon. (\subref{fig:arch}) \textit{System architecture}. Number of vLLM instances depends on task load. Load balancer dispatches to the least utilized instance.}\label{fig:net}
\end{figure}
\begin{algorithm}[t]
\caption{Simulation protocol}\label{alg:protocol}
\begin{algorithmic}[1]
    \Require {\(x(t) \in \Set{X}\) and \(G = (V,E)\)}
    \State {Initialize post \(p_i\) with opinion \(x_i(t)\) using \emph{encoding} prompt for all agents \(i \in V\).}
    \ForEach {agent \(i \in V\)}
        \State \(\Nset(i,t) \gets \{ j : i \sim j \land |x_i(t) - x_j(t)| \leq \epsilon \}\)
        \State Concatenate all posts \(p_j\) for \(j \in \Nset(i,t)\) to form a timeline.
        \State Update textual opinion \(y(t)\) from timeline using \emph{dynamics} prompt.
        \State Decode \(y(t)\) to scalar opinion \(x_i(t+1)\) using \emph{decoding} prompt.
    \EndFor
    \State \Return \(x(t+1)\)
\end{algorithmic}
\end{algorithm}
The full simulation protocol for a single transition \(\vx(t) \mapsto \vx(t+1)\) is shown in \cref{alg:protocol}. Computationally, an important aspect is that each agent needs only access to their ego network in \(G\) to perform their individual update. Hence, agent updates can be performed in parallel once the initial opinion posts are available (which can be computed in parallel too).
%
Nevertheless, LLM-ABM simulations can be time-consuming due to the large number of calls to an LMM with significant computational demands. Every action within an LLM agent's behavior can require a sequence of LLM inference calls whose number increases with the agent count (N) in the simulation. Due to each agent's dual role in initiating actions and responding to others, it can potentially escalate complexity beyond linear. More agents also extend the prompt sequence, considering more observations when generating reactions. Standard transformer implementations~\citep{Vaswani2017-ld} have a quadratic complexity with sequence length.
These LLM-ABM's requirements can exceed the usual rate limits of commercial API-based LLMs, making direct use of such services impractical. Open-source models like \llama-2 70B, when used without parallelism, also exhibit prolonged response generation times. To tackle these challenges, we develop an end to end LLM-ABM framework. Our approach encompasses two primary components: an optimized agent implementation and an efficient LLM inference server. We will be releasing the code associated with this paper.
In the agent implementation, we distinguish between tasks requiring simultaneous execution and those amenable to parallelization. We also use caching for repeated tasks such as initial persona summarization to expedite startup time and minimize unnecessary LLM calls. We utilize LangChain as the interface for LLM interaction, ensuring compatibility with various LLMs models, including \llama.
For the LLM backend, we implement a load balancer with vLLM inference serving engines~\citep{Kwon2023-fb}. Multiple vLLM instances can run on different nodes, with an NGINX-based load balancer managing traffic and automatic failure handling.
Using these optimizations, our system can facilitate large simulations, handling a pairwise conversation with 5000 agents in an hour when running with 16 nodes (with 8 V100 GPUs each). The majority of the experiments detailed in this paper are therefore completed within minutes. However, it is important to note that some LLM constraints can still impact scalability. For instance, the 4096 context size of \llama-2 imposes a constraint on the amount of text that a single agent can process, causing the number of neighbors used in the simulation to be limited by content length, particularly when aggregating interactions from other agents. For LLMs with larger context length, these limitations disappear in our implementation.

\section{Experiments}\label{sec:results}
In the following, we evaluate the ability of LLMs to reproduce the dynamics of the DeGroot and Hegselmann-Krause models using the framework of \cref{sec:method}. Our goal is to evaluate in what way LLMs capture the dynamics we would expect, which prompting techniques are (un)successful, and how sensitive they are to the choice of prompts.
For this purpose, we run simulations of opinion dynamics related to \emph{free market versus a planned economy}. The advantage of this topic is that it allows us to state different view points in terms of  positive and negative framings, i.e., an opinion pro free market corresponds to an opinion contra planned economy, and vice versa. This will allow us in our experiments to investigate possible biases of LLMs against negative feedback in general (which might, for instance, be introduced through alignment methods). Since both framings are semantically close to identical, this will allow us also to analyze this in terms of sensitivity and reliability.
To encode opinion values, we are mapping manually from the range of \([-5, 5]\) to an 11-point text opinion scale in our main experiments. In this mapping, the sign is interpreted as supporting one of the two contrarian positions (pro free market versus pro planned economy), and the magnitude is interpreted as the strength of support, mapping \([1, 5]\) to \textit{a bit in favor of}, \textit{slightly in favor of}, \textit{moderately in favor of}, \textit{clearly in favor of}, or \textit{strongly supportive of} accordingly. 
Magnitude \(0\) is used to represent \emph{indifference} or preference for a mixed approach.  
In addition to this manual mapping, we include an ablation where the LLM is tasked to convert the numeric opinion autonomously into text, i.e., without a manual mapping, in the encode-decode consistency section.
As loss functions in \cref{eq:mcons}, we will use the absolute distance between reference and LLM dynamics as well as the indicator function whether the sign of opinions (or their polarity) agree. Formally, these loss functions are defined as:
\begin{align*}
    \loss_\text{Distance}(f(x), h(x)) = |f(x) - h(x)|, \quad\text{and}\quad
    \loss_\text{Polarity}(f(x), h(x)) = \mathbb{1}[\operatorname{sgn} \lceil f(x) \rfloor \neq \operatorname{sgn} h(x)] .
\end{align*}
where \(\lceil x \rfloor \) denotes rounding of $x \in \R$ to the nearest integer.



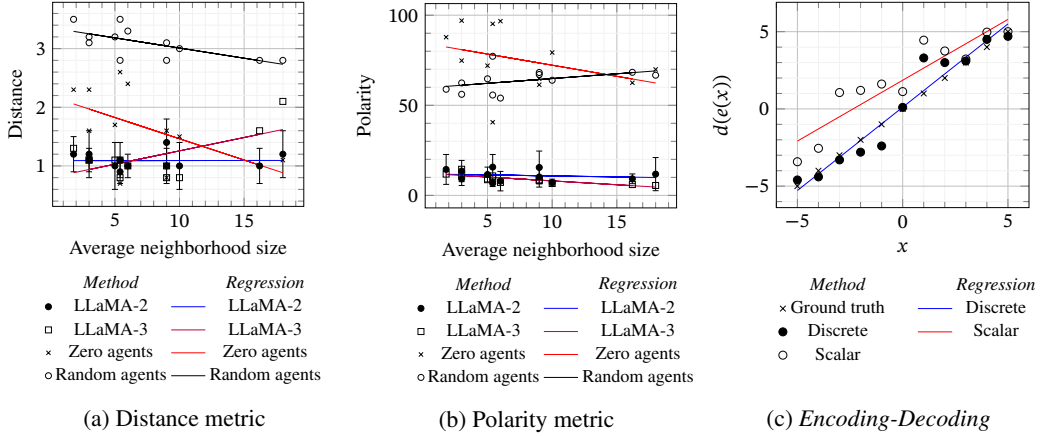
\begin{figure}
    \begin{subfigure}[T]{.33\linewidth}
        \resizebox{.95\linewidth}{!}{%
        \centering
        \begin{tikzpicture}
        \begin{axis}[%
            width=6cm,
            xlabel={Average neighborhood size},
            ylabel={Distance},
            grid style={line width=.1pt,draw=gray!10},
            major grid style={line width=.1pt,draw=gray!50},
            minor tick num=4,
            grid=both,
            legend columns=2,
            legend style={draw=none, font=\small, anchor=north, at={(0.5,-0.3)}},
            mark size=1.5,
            mark options={black},
        ]
            \addlegendimage{empty legend}
                \addlegendentry{\textit{Method}};
            \addlegendimage{empty legend}
                \addlegendentry{\textit{Regression}};
            \addplot [only marks, error bars/.cd, y dir=both, y explicit] table [col sep=comma, x={Average aggragated neighbours}, y={LLM Agents}, y error={LLM Agents error margin}] {./dat/distance_plot.csv};
                \addlegendentry{LLaMA-2};
            \addplot [no marks, blue] table [col sep=comma, x={Average aggragated neighbours}, y={create col/linear regression={y={LLM Agents}}}] {./dat/distance_plot.csv};
                \addlegendentry{LLaMA-2};
            \addplot [only marks, mark=square] table [col sep=comma, x={Average aggragated neighbours}, y={LLM Agents 3}] {./dat/distance_llama3.csv};
                \addlegendentry{LLaMA-3};
            \addplot [no marks, purple] table [col sep=comma, x={Average aggragated neighbours}, y={create col/linear regression={y={LLM Agents 3}}}] {./dat/distance_llama3.csv};
                \addlegendentry{LLaMA-3};
            \addplot[only marks, mark=x] table [col sep=comma, x={Average aggragated neighbours}, y={Zero Agents}] {./dat/distance_plot.csv};
                \addlegendentry{Zero agents};
            \addplot[no marks, red] table [col sep=comma, x={Average aggragated neighbours}, y={create col/linear regression={y={Zero Agents}}}] {./dat/distance_plot.csv};
                \addlegendentry{Zero agents};
            \addplot[only marks, mark=o] table [col sep=comma, x={Average aggragated neighbours}, y={Random Agents}] {./dat/distance_plot.csv};
                \addlegendentry{Random agents};
            \addplot[no marks, black] table [col sep=comma, x={Average aggragated neighbours}, y={create col/linear regression={y={Random Agents}}}] {./dat/distance_plot.csv};
                \addlegendentry{Random agents};
        \end{axis}
        \end{tikzpicture}}
        \caption{Distance metric}\label{fig:distance}
    \end{subfigure}%
    \begin{subfigure}[T]{.34\linewidth}
        \centering
        \resizebox{.99\linewidth}{!}{%
        \begin{tikzpicture}
        \begin{axis}[%
            width=6cm,
            xlabel={Average neighborhood size},
            ylabel={Polarity},
            grid style={line width=.1pt,draw=gray!10},
            major grid style={line width=.1pt,draw=gray!50},
            minor tick num=4,
            grid=both,
            legend columns=2
            legend cell align={left},
            legend style={draw=none, font=\small, anchor=north, at={(0.5,-0.3)}},
            mark size=1.5,
            mark options={black},
        ]
            \addlegendimage{empty legend}
                \addlegendentry{\textit{Method}};
            \addlegendimage{empty legend}
                \addlegendentry{\textit{Regression}};
            \addplot [only marks, error bars/.cd, y dir=both, y explicit] table [col sep=comma, x={Average aggragated neighbours}, y=LLM_Agents, y error={LLM_Agents_errors}] {./dat/polarity_plot.csv};
                \addlegendentry{LLaMA-2}
            \addplot [no marks, blue] table [col sep=comma, x={Average aggragated neighbours}, y={create col/linear regression={y={LLM_Agents}}},] {./dat/polarity_plot.csv};
                \addlegendentry{LLaMA-2}
            \addplot [only marks, mark=square] table [col sep=comma, x={Average aggragated neighbours}, y=LLM_Agents_3] {./dat/polarity_llama3.csv};
                \addlegendentry{LLaMA-3}
            \addplot [no marks, purple] table [col sep=comma, x={Average aggragated neighbours}, y={create col/linear regression={y={LLM_Agents_3}}},] {./dat/polarity_llama3.csv};
                \addlegendentry{LLaMA-3}
            \addplot [only marks,mark=x] table [col sep=comma, x={Average aggragated neighbours}, y={Zero_Agents}] {./dat/polarity_plot.csv};
                \addlegendentry{Zero agents}
            \addplot [red,no marks] table [col sep=comma, x={Average aggragated neighbours}, y={create col/linear regression={y={Zero_Agents}}}] {./dat/polarity_plot.csv};
                \addlegendentry{Zero agents}
            \addplot [only marks,mark=o] table [col sep=comma, x={Average aggragated neighbours}, y={Random_Agents}] {./dat/polarity_plot.csv};
                \addlegendentry{Random agents}
            \addplot [black,no marks] table [col sep=comma, x={Average aggragated neighbours}, y={create col/linear regression={y={Random_Agents}}}] {./dat/polarity_plot.csv};
                \addlegendentry{Random agents}
        \end{axis}
        \end{tikzpicture}}
        \caption{Polarity metric}\label{fig:polarity}
    \end{subfigure}%
    \begin{subfigure}[T]{.33\linewidth}
    \pgfplotstableread[row sep=crcr]{
    x y label \\
    5 4.7 a \\
    4 4.5 a \\
    3 3.1 a \\
    2 3.0 a \\
    1 3.3 a \\
    0 0.1 a \\
    -1 -2.4 a \\
    -2 -2.8 a \\
    -3 -3.3 a \\
    -4 -4.4 a \\
    -5 -4.6 a \\
    }\datdiscrete
    \pgfplotstableread[row sep=crcr]{
    x y label \\
    5 5 b \\
    4 4.97 b \\
    3 3.23 b \\
    2 3.75 b \\
    1 4.45 b \\
    0 1.12 b \\
    -1 1.61 b \\
    -2 1.2 b \\
    -3 1.06 b \\
    -4 -2.55 b \\
    -5 -3.42 b \\
    }\datscalar
    \pgfplotstableread[row sep=crcr]{
    x y label \\
    5 5 c \\
    4 4 c \\
    3 3 c \\
    2 2 c \\
    1 1 c \\
    0 0 c \\
    -1 -1 c \\
    -2 -2 c \\
    -3 -3 c \\
    -4 -4 c \\
    -5 -5 c \\
    }\dattrue
    \centering
    \resizebox{.98\linewidth}{!}{%
    \begin{tikzpicture}
        \begin{axis}[%
            width=6cm,
            xlabel={$x$},
            ylabel={$d(e(x))$},
            minor tick num=4,
            grid style={line width=.1pt,draw=gray!10},
            major grid style={line width=.1pt,draw=gray!50},
            grid=both,
            legend columns=2,
            legend style={draw=none, font=\small, anchor=north, at={(0.5,-0.3)},/tikz/every even column/.append style={column sep=0.5cm}},
            ]
            \addlegendimage{empty legend}
                \addlegendentry{\textit{Method}};
            \addlegendimage{empty legend}
                \addlegendentry{\textit{Regression}};
            \addplot[only marks, mark=x] table {\dattrue};
                \addlegendentry{Ground truth}
            \addplot[blue,no marks] table[y={create col/linear regression={y=y}}] {\datdiscrete};
                \addlegendentry{Discrete}
            \addplot[only marks, mark=*] table {\datdiscrete};
                \addlegendentry{Discrete}
            \addplot[red,no marks] table[y={create col/linear regression={y=y}}] {\datscalar};
                \addlegendentry{Scalar}
            \addplot[only marks, mark=o] table {\datscalar};
                \addlegendentry{Scalar}
        \end{axis}
    \end{tikzpicture}}
    \vspace{8pt}
    \caption{\textit{Encoding-Decoding}}\label{fig:encdec}
    \end{subfigure}
\caption{\textit{Consistency Evaluation}}\label{fig:main}
\end{figure}

\paragraph{Model consistency}
To meaningfully evaluate the sensitivity of an LLM-ABM to prompt variations, we first need to establish that a base model \(\theta\) is sufficiently consistent with the reference model via \cref{def:mcons}. For this purpose, we first initialize opinions uniformly at random within \([-1, 1]\) for the entire population of \(N\) agents. Subsequently, we run the Hegselmann-Krause model for a total of 20 epochs. At each time step, the updates are computed according to \cref{alg:protocol}, where the input state \(x(t)\) of the reference model at each step is subsequently fed into the LLM agents to compute the simulation error. 
The set of prompts \(\theta_\text{enc}\), \(\theta_\text{dec}\), and \(\theta_\text{dyn}\) used for this process can be found in \cref{tab:prompts} in appendix \ref{app:prompt_example}.
%
%
%
We evaluate distance and polarity, using \(N = 20\) agents compared to the reference model. Distance metrics are computed by rescaling the reference model's opinions from a continuous \(([-1, 1])\) scale to the LLMs \(([-5, 5])\) scale. Polarity, a measure of opinion divergence, rounds values within \(x(t) \in [-0.5, 0.5]\) to 0 and counts sign mismatches. \Cref{fig:polarity} presents the polarity percentage of total updates. In both figures, the LLM agents value are shown with its error bar representing a \(0.95\) confidence interval.
To introduce a comparison scale, we report two baselines. (i) A \emph{random agent}, i.e., an agent whose opinions are sampled from a discrete uniform distribution over integers such that \(\vx(t)_i = \mathcal{U}\{-5, 5\}\). (ii) A \emph{zero agent}, which consistently selects zero as its opinion, serves a dual purpose: it represents an agent who always chooses `indifferent' and an agent who consistently chooses the network's average opinion as the network's initial opinion is sampled from a Gaussian distribution centered at 0. The full details of the experiments including runs with \(N = 50\) can be found in \cref{appendix:distance}.
\Cref{fig:distance,fig:polarity} shows the results of these experiments. It can be seen that the base model provides a good approximation of the reference dynamics (e.g., polarity is inaccurate in less than 10\% cases) and provides substantial improvements over the comparison models. Hence, this base model \(\theta\) is appropriate to further investigate the sensitivity of the LLM-ABM.
\begin{table}[b]
\caption{(\subref{table_sensitivity_to_prompts}) \emph{Prompt sensitivity} and (\subref{tab:negbias}) \emph{Negativitiy bias} for HK simulations using \llama-2. Examples of differences between base model and variants are highlighted in green (base), red (variant).}\label{tab:res}
\begin{subfigure}{\linewidth}
\begin{minipage}[b]{.05\linewidth}
    \caption{}\label{table_sensitivity_to_prompts}
    \vspace{2em}
\end{minipage}%
\begin{minipage}[b]{.95\linewidth}
\small
\begin{tabular}{lllllp{2pt}llp{2pt}lll}
\toprule
\multicolumn{3}{l}{} & \multicolumn{2}{c}{\textbf{Prompt \(\theta_\text{base}\)}} & & \multicolumn{2}{c}{\textbf{Prompt \(\theta_\text{wording}\)}} & & \multicolumn{2}{c}{\textbf{Prompt \(\theta_\text{logic}\)}} \\
\cmidrule(r){4-5}\cmidrule(l){7-8}\cmidrule(l){10-11}
\(N\) & \(p\) & \(\epsilon\) & Distance & Polarity \% & & Distance & Polarity \% & & Distance & Polarity \%\\
\midrule
20 & 0.3 & 0.9 & 1.1 (0.3) & 15.7 (7.0)  & & 1.5 (0.2) & 26.3 (19.4) & & 2.5 (0.2) & 58.0 (26.5) \\
20 & 0.5 & 0.9 & 1.4 (0.4) & 15.5 (9.1)  & & 1.6 (0.3) & 18.4 (16.3) & & 2.3 (0.4) & 55.8 (31.7) \\
20 & 0.9 & 0.9 & \goodcell 1.0 (0.3) & \phantom{1}9.2 (2.7)  & & \badcell 1.7 (0.3) & 23.2 (18.8) & & \badcell 2.3 (0.3) & 48.4 (30.3) \\
20 & 1.0   & 0.9 & 1.2 (0.4) & \goodcell 11.8 (9.2)  & & 1.7 (0.2) & \badcell 44.0 (22.9) & & 2.2 (0.4) & \badcell 49.3 (33.0) \\
\bottomrule
\end{tabular}
\end{minipage}
\end{subfigure}

\vspace*{1ex}

\begin{subfigure}{\linewidth}
%
\begin{minipage}[b]{.05\linewidth}
    \caption{}\label{tab:negbias}
    \vspace{1.5em}
\end{minipage}%
\begin{minipage}[b]{.95\linewidth}
\small
\begin{tabular}{llccc}
\toprule
& & \multicolumn{3}{c}{\textbf{Author}} \\
\cmidrule{3-5}
& & Pro Free Market & Pro Planned Economy & Contra Free Market\\
\midrule
\multirow{3}{*}{{\textbf{Reader}}} & Pro Free Market     & \goodcell \phantom{-}1.1 (0.0)          &  -1.0 (0.0) & -0.9 (0.0)\\
& Pro Planned economy & -1.0 (0.0)& \goodcell \phantom{-}1.1 (0.0)& -   \\ 
& Contra Free Market & -1.0 (0.0)& - & \badcell\phantom{-}0.8 (0.1) \\
\bottomrule
\end{tabular}
\end{minipage}
\end{subfigure}
\end{table}

\paragraph{Encode-decode consistency}\label{sec:encdec-cons}
To evaluate the encoding-decoding consistency as in \cref{def:edcons}, we repeatedly encode and decode opinion values within \([-5,5]\) and record the mean consistency over all repetitions. We compare two pairs of encoding-decoding prompts (i) \emph{discrete}, where we manually discretize the numeric opinions into the textual range as discussed before and (ii) \emph{scalar}, where the LLM receives the raw numerical opinion and is tasked to convert this to a textual representation autonomously. The results of these experiments are shown in \cref{fig:encdec}. It can be seen that the encoding-decoding scheme can have substantial impact and is far from trivial. In general, the discrete scheme provides clearly better results than the scalar scheme, and is reasonably aligned with the ground truth. However, for opinions close to zero, the codec quality deteriorates which indicates challenges for the LLM to process nuances close to neutral stances.


\paragraph{Sensitivity to prompts}
As seen in the previous experiments, it is possible to find instruction prompts that provide a reasonable approximation to the reference dynamics. However, consistency is only a necessary and not a sufficient condition for an LLM-ABM to be useful. A second necessary condition is reliability, which we analyze in the following. For this purpose, we create  two prompt sets that are semantically similar to the original prompts \(\theta_\text{base}\) but have two main differences:
\begingroup
\advance\leftmargini -2em
\begin{quote}
\begin{description}
    \item[Prompt wording \(\theta_\text{wording}\):] In this prompt set, the wording of the decoding prompt is slightly changed. In \(\theta_\text{base}\), the agent is asked if the polarity of a post equals \(1\) and selecting \(-1\) otherwise, while \(\theta_\text{wording}\), asks whether the polarity equals \(1\) or \(-1\). Clearly, both prompts are semantically identical. All other prompts are unchanged.
    \item[Prompt logic \(\theta_\text{logic}\):] In this prompt set, the decoding prompt \(\theta_\text{dec}\) is changed such that it directly asks for opinion score on a scale of  \([-5, 5]\), rather than first asking for the polarity \(\{-1, 0, 1\}\) and then the magnitude of the opinion \([1,5]\) as in \(\theta_\text{base}\). Again, both prompts are semantically identical and \(\theta_\text{enc}\), \(\theta_\text{dyn}\) remain unchanged. 
    \item[Prompt newline \(\theta_\text{\textbackslash n}\):] In this prompt set, the only change compared to \(\theta_\text{base}\) is the removal of newline characters \(\textbackslash n\)  and apostrophe ' in all prompts.
\end{description}
\end{quote}
\endgroup
The results for \(\theta_\text{wording}\) and \(\theta_\text{logic}\) are shown in \cref{table_sensitivity_to_prompts}. It can be seen that the choice of prompts can have a significant impact on the consistency of the LLM-ABM. For instance, the percentage of posts in which the polarity is wrongly assessed jumps from 11.8\% on average to 44.0\% and 49.3\%, a 4x increase and worse than a random agent. Similar results can be observed for the distance metric  in \cref{table_sensitivity_to_prompts}, as well as for \(\theta_\text{\textbackslash n}\) in table \ref{appendix:theta_backslash_distance} \cref{appendix:theta_backslash}, where the distance metrics drops from \(2.1\) to \(1.2\) for \textsc{LLaMA-3}. An important observation is also that these results are not dependent on LLM quality, since \textsc{LLaMA}-3 is not necessarily more robust than \textsc{LLaMA}-2. Overall, these results demonstrate well the limited reliability of LLM-ABMs with regard to the choice of prompts.

\paragraph{Negativity bias}
Finally, we also evaluate the impact of question framing on the LLM's opinion generation. We compare the performance of two prompts, one contrasting two positively framed stances (pro free-market versus pro planned economy) and one contrasting a positive and negative stance (pro free-market versus contra free-market). We design a simplified scenario where an \emph{author} agent is tasked with generating a post that expresses its opinion. Subsequently, a \emph{reader} agent is directed to express their opinion about the author's post. Opinion about topics as well as opinions about posts are on a scale ranging from [\textit{strong disagree}, \textit{disagree}, \textit{indifferent}, \textit{agree}, \textit{strong agree}], corresponding to \([-2, 2]\). Each reader considers their own opinion about the topics (which is encoded in their persona) when assessing a post. We show the average opinion for each combination of these framings in \cref{tab:negbias}. 
It can be seen that there exist again clear differences in the consistency where the positive-only framing shows better results. Since both prompts are semantically very close, this supports also the sensitivity of LLM-ABMs and the associated reliability issues.

\section{Related work}\label{sec:related}

LLM-based agent-based models, denoted LLM-ABM, have recently become feasible due to the ongoing development of large language models. Like classical ABMs, the applications are wide-ranging and diverse, including virtual-world simulation, social media and opinion dynamics models, and analysis of physical systems.~\citet{Park2023-ur} introduced generative agents simulating the behavior of a small town displaying unexpected emergent behavior.~\citet{Kaiya2023-wt} introduced a virtual society simulation of agents in a 3D environment.~\citet{Gao2023-fm} and \citet{Park2023-ur} consider social media simulations of large numbers of agents.~\cite{Tornberg2023-bt} simulated social media users as agents and performed synthetic experiments evaluating the effects of different ranking algorithms determining which posts the agents observe.
\cite{Chuang2023-uq} considered LLM-driven opinion dynamics simulations similar to those examined here.  This past research on social simulation and opinion dynamics via LLM-ABM demonstrated that rich, complex behavior could be created, but calibrating the dynamics was largely considered crucial future research.  Here we provide a framework to evaluate consistency with known theories of opinion dynamics, grounding LLM-ABM dynamics with respect to a reference model. Furthermore, we extend the research focus to the sensitivity of LLM-ABMs to changes in their micro-specifications and their suitability for theory-building.
While not considering an explicit reference,~\citet{Taubenfeld2024-zu} discussed limitations of LLM-ABM's for simulating political debates, finding systematic intrinsic biases towards moderation that led to debaters' opinions shifting to common ground. Another approach to calibrated behavior is to restrict LLM dynamics to match a known parameterized model, and \citet{Memery2023-ya} examined how well LLM's can perform parameter inference in physical systems.  

\section{Discussion}\label{sec:discussion}
In this work, we have proposed a new evaluation framework to understand the ability of LLMs to replace ABMs for the simulation of social dynamics. By treating LLMs as black-box function and grounding their dynamics in the dynamics of references models, we were able to show that the simulated dynamics of LLMs can vary substantially relative to semantically minor changes in prompts. While LLMs are able to capture the intended dynamics with sufficient prompt engineering, it is therefore difficult to get reliable insights from the prompts that parameterize the simulation. This puts into question their use as ABMs, at least in their current form. If minor, semantically equivalent changes can have substantial impact on a simulation's outcome, no insights can be gained for theory-building and analysis by analyzing the microbehavior encoded in the prompts.

While our results already provide important insights, they are also limited in two aspects that point to promising directions for future work. First, the scale of our experiments is currently limited by the context length that is achievable by models such as \textsc{LLaMA}-2. Since scale and corresponding network topologies can have non-trivial effects on social dynamics, overcoming this scaling limitation can provide for insights into the capabilities of LLMs. Second, our results are currently attained via manual search over prompt-space. While this allows to control semantic similarity in the prompt search, it does not scale to a large-scale investigation of the prompt space. Automated prompt optimization methods such as \textsc{PromptBreeder}~\citep{Fernando2023PromptbreederSS} and \textsc{AdvPrompter}~\citep{Paulus2024AdvPrompterFA} are promising directions to overcome this limitation.

\bibliography{references}
\bibliographystyle{colm2024_conference}
\clearpage
\appendix
\section{Supplementary Material}
\subsection{Prompt Examples}
\label{app:prompt_example}
\begin{table}[h]
    \centering
    \small
    \caption{\textit{Prompt examples.}}\label{tab:prompts}
    \begin{tabular}{p{.45\linewidth}p{.45\linewidth}}
        \toprule
        \textbf{Encoding prompt}
        & \textbf{Dynamics prompt} \\
        \cmidrule(r){1-1} \cmidrule(l){2-2}
        \ttfamily
        \{persona\}\newline
        You are \{opinion\} \{topic\}.\newline
        Compose a tweet about \{topic\} to reflect your stance from your personal experience. Please rationalize the link between your personal experience and your stance in your tweet. Remember, there's no need to be overly polite --- this is your personal tweet. ONLY reply with the tweet and nothing else.\newline
        Do not add comments nor explanation.
        &
        \ttfamily
        \{persona\}\newline
        Read tweets from your network:\newline`\{timeline\}'\newline
        After analyzing these posts, it's time to formulate your stance. Your options are `completely neutral towards', `a bit in favor of', `slightly in favor of', `moderately in favor of', `clearly in favor of', or `strongly supportive of'. Your opinion should be a balanced reflection of the sentiments expressed in the tweets you've analyzed, with each tweet given equal consideration.\newline
        Keep it candid and concise. Provide your opinion regarding the planned economy versus free market economy debate. Your response should include your opinion and specific references to the posts that influenced your stance.\\
        \midrule
        \multicolumn{2}{l}{\textbf{Decoding prompt}}  \\
        \midrule
        \multicolumn{2}{p{.9\columnwidth}}{%
        \ttfamily
        Please read the following text:\newline \{text\}\newline Based on the author's perspective, do they support \{topic\}? Reply with 'yes' or 'no'. Don't add anything else. \newline\newline
        Please carefully read the following text: \{text\} \newline 
        Based on the author's perspective, evaluate their level of support for \{topic\}. Choose from the following options: 'completely neutral towards', 'a bit in favor of', 'slightly in favor of', 'moderately in favor of', 'clearly in favor of', or 'strongly supportive of'. If the author prefers a mixed approach, regardless of the intensity of their stance, please select 'completely neutral towards'. Ensure your response is limited to these options and does not include any additional information.}\\
        \midrule
        \multicolumn{2}{l}{\textbf{Persona}}  \\
        \midrule
        \multicolumn{2}{p{.9\columnwidth}}{%
        \ttfamily
        You like to remodel homes.\newline
        You like to go hunting. \newline
        You like to shoot a bow. \newline
        Your favorite holiday is Halloween.}\\
        \bottomrule
    \end{tabular}
\end{table}
\begin{table}[h]
    \centering
    \small
    \caption{\textit{Prompt \(\theta\) examples.}}\label{tab:prompts_theta}
    \begin{tabular}{p{.45\linewidth}p{.45\linewidth}}
        \toprule
        \multicolumn{2}{l}{\textbf{Decoding prompt \(\theta_\text{wording}\)}}  \\
        \midrule
        \multicolumn{2}{p{.9\columnwidth}}{%
        \ttfamily
        Please read the following text:\newline \{text\} \newline
        Based on the author's perspective, determine which economic system they support a more planned economy or a free market economy? Please choose your answer between 'planned economy' or 'free market economy'. Only choose between this two and do not include any additional information in your response. Again: reply with 'planned economy' or 'free market economy' and nothing else.\newline
        Please carefully read the following text: \newline \{text\} \newline 
        Based on the author's perspective, evaluate their level of support for \{topic\}. Choose from the following options: 'completely neutral towards', 'a bit in favor of', 'slightly in favor of', 'moderately in favor of', 'clearly in favor of', or 'strongly supportive of'. If the author prefers a mixed approach, regardless of the intensity of their stance, please select 'completely neutral towards'. Ensure your response is limited to these options and does not include any additional information.
        }\\
        \midrule
        \multicolumn{2}{l}{\textbf{Decoding prompt logic \(\theta_\text{logic}\)}}  \\
        \midrule
        \multicolumn{2}{p{.9\columnwidth}}{%
        \ttfamily
        Please read the following text: \newline \{text\} \newline
        Based on the author's perspective, evaluate their level of support for Free market economy. Choose from the following options: 'strongly opposed to', 'clearly against', 'moderately against', 'slightly against', 'a bit against', 'completely neutral towards', 'a bit in favor of', 'slightly in favor of', 'moderately in favor of', 'clearly in favor of', to 'strongly supportive of'. If the author prefers a mixed approach, regardless of the intensity of their stance, please select 'completely neutral towards'. Ensure your response is limited to these options and does not include any additional information. Remember, if the author is supportive to planned economy, it implies that the author is opposed to free market economy, and vice versa. Please return author's opinion specifically on the concept of a Free market economy. Only reply and given option and nothing else."
        }\\
        \bottomrule
    \end{tabular}
\end{table}

\clearpage

\subsection{Example of feed content and subsequent update of agent’s opinion}
\begin{table}[h]
    \small
    \caption{Example of feed content and subsequent update of agent's opinion.}
    \begin{tabular}{lp{.77\linewidth}}
        \toprule
        \multicolumn{2}{l}{\textbf{Feed}}\\
        \midrule
        \texttt{@rowan} & After struggling with inconsistent employment and financial instability, I've come to realize the importance of a planned economy. It's time for the government to step in and create a system that ensures everyone has access to basic needs like healthcare, education, and a living wage. Let's break free from the exploitative grip of capitalism and build a society that values people over profit. \#PlannedEconomy \#Socialism \#EconomicJustice \\
        \cmidrule(r){1-1}\cmidrule(l){2-2}
        \texttt{@jakob} & As someone who\'s worked in construction and seen the benefits of organized labor, I think a planned economy has its perks. But as a small business owner, I also know the importance of flexibility and innovation. Ultimately, a mix of both systems works best for me. \#plannedconomy \#freeMarket \#pragmatism \\
        \cmidrule(r){1-1}\cmidrule(l){2-2}
        \texttt{@mallory} & As someone who's lived in both planned and free-market economies, I can say that neither system is perfect. But, I do believe that a mixed economy with a strong social safety net is the way to go. It's important to have some regulation to prevent exploitation, but also allow for innovation and individual freedom. \#PlannedEconomy \#FreeMarketEconomy \#MixedEconomy" \\
        \cmidrule(r){1-1}\cmidrule(l){2-2}
        \texttt{@abril} & As someone who\'s had to juggle multiple jobs and a tight budget while pursuing a degree, I\'ve come to realize the importance of a planned economy. It\'s not about stifling innovation, but about creating a safety net for those who need it most. \#PlannedEconomy \#SocialSafetyNet \#PersonalExperience.\\
        \midrule
        \textbf{Agent Update} \\
        \midrule
        \multicolumn{2}{p{.945\columnwidth}}{%
        After analyzing the given tweets, I have formed an opinion that leans slightly in favor of a planned economy.\vspace{1em}}\\
        \multicolumn{2}{p{.945\columnwidth}}{%
        While I acknowledge the importance of innovation and individual freedom in a free market economy, I believe that a planned economy can provide a necessary safety net for those who need it most, as expressed by \texttt{@abril} \texttt{@jakob}'s tweet highlights the benefits of organized labor in a planned economy, which resonates with my opinion. Moreover, \texttt{@mallory}'s view that a mixed economy with a strong social safety net is the best approach aligns with my stance. However, I also recognize the potential drawbacks of a planned economy, such as stifling innovation, as mentioned by \texttt{@rowan}. Therefore, I believe that a balanced approach that incorporates elements of both systems is the most effective way forward.\vspace{1em}}\\
        \multicolumn{2}{p{.945\columnwidth}}{%
        In conclusion, my opinion is slightly in favor of a planned economy, as I believe it can provide essential support for those in need while still allowing for innovation and individual freedom.
        }\\
        \bottomrule
    \end{tabular}
\end{table}
\clearpage
\subsection{Comparative analysis of model consistency}
\begin{table}[h]
\label{appendix:distance}
\caption{Distance from the Reference Model. In this context, N represents the number of agents involved in the simulation, p denotes the connection probability of the network, and $\epsilon$ is defined as per the Hegselmann-Krause models. The deviation is quantified as the absolute difference from the reference model. The reference model is multiplied by 5 to transition from the [-1, 1] scale to align with the [-5, 5] scale used by the LLM Agents. The 95\% confidence interval width is in parentheses. Note that \llama-3 permits larger neighborhood size due to its 8192 context size.}
\small
\begin{center}
\begin{tabular}{ccccccc}
\toprule
\(N\) & \(p\) & \(\epsilon\) & \llama-2 & \llama-3 & Zero Agents & Random Agents \\
\midrule
20 & 0.3 & 0.3 & 1.2 (0.3) & 1.3 (0.2) & 2.3 (0.5) & 3.5 (0.2) \\
20 & 0.3 & 0.5 & 1.2 (0.4) & 1.1 (0.2) & 1.6 (0.7) & 3.1 (0.2) \\
20 & 0.3 & 0.9 & 1.1 (0.3) & 0.8 (0.4) & 0.7 (0.4) & 2.9 (0.1) \\
20 & 0.5 & 0.3 & 1.1 (0.2) & 1.1 (0.3) & 2.3 (0.4) & 3.4 (0.2) \\
20 & 0.5 & 0.5 & 1.0 (0.4) & 1.1 (0.4) & 1.7 (0.8) & 3.1 (0.3) \\
20 & 0.5 & 0.9 & 1.4 (0.4) & 0.8 (0.4) & 0.8 (0.4) & 2.9 (0.1) \\
20 & 0.9 & 0.3 & 0.9 (0.2) & 1.1 (0.2) & 2.6 (0.3) & 3.4 (0.2) \\
20 & 0.9 & 0.5 & 1.0 (0.3) & 1.0 (0.4) & 1.6 (0.7) & 3.1 (0.3) \\
20 & 0.9 & 0.9 & 1.0 (0.3) & 1.6 (0.5) & 1.0 (0.4) & 2.8 (0.1) \\
20 & 1.0 & 0.3 & 1.0 (0.2) & 1.0 (0.3) & 2.4 (0.3) & 3.3 (0.1) \\
20 & 1.0 & 0.5 & 1.0 (0.4) & 0.8 (0.4) & 1.5 (0.7) & 3.0 (0.2) \\
20 & 1.0 & 0.9 & 1.2 (0.4) & 2.1 (0.4) & 1.1 (0.5) & 3.0 (0.1) \\
50 & 0.3 & 0.3 & 0.9 (0.2) & 1.0 (0.2) & 2.1 (0.5) & 3.3 (0.2) \\
50 & 0.3 & 0.5 & 1.1 (0.3) & 1.1 (0.5) & 1.1 (0.5) & 2.9 (0.1) \\
50 & 0.3 & 0.9 & 1.2 (0.5) & 2.0 (0.4) & 0.4 (0.2) & 2.8 (0.0) \\
50 & 0.5 & 0.3 & - & 1.1 (0.3) & 2.2 (0.3) & 3.3 (0.2)\\
50 & 0.5 & 0.5 & - & 0.9 (0.3) & 1.0 (0.4) & 2.9 (0.1) \\
50 & 0.9 & 0.3 & - & 0.7 (0.2) & 2.4 (0.2) & 3.3 (0.1) \\
50 & 1.0 & 0.3 & - & 0.8 (0.2) & 2.5 (0.2) & 3.4 (0.1) \\
\bottomrule
\end{tabular}
\end{center}
\end{table}

\begin{table}[h]
\label{appendix:polarity}
\caption{Polarity percentage of agents. Given the continuous scale of the reference model and the discrete scale of the LLM agents, direct polarity comparison is complex. To address this, we report polarity \% when agents' opinions diverge from the reference model. After scaling the reference model by a factor of 5 to match the [-5, 5] range and rounding small value opinions within -0.5 to 0.5 to 0, polarity count is defined as a mismatch in opinions value signs. We then report its percentage out of total updates.}
\small
\begin{center}
\begin{tabular}{ccccccc}
\toprule
\(N\) & \(p\) & \(\epsilon\) & \llama-2 & \llama-3 & Zero Agents & Random Agents \\
\midrule
20 & 0.3 & 0.3 & 14.4 (8.3) & 11.8 (2.0) & 87.8 (9.2)  & 58.7 (3.7) \\
20 & 0.3 & 0.5 & 13.5 (5.9) & 14.5 (3.0) & 74.8 (16.5) & 63.9 (6.6) \\
20 & 0.3 & 0.9 & 15.7 (7.0) & 10.8 (2.2) & 40.6 (17.7) & 75.7 (6.8) \\
20 & 0.5 & 0.3 & 9.2 (3.7)  & 11.2 (3.9) & 97.0 (3.1)  & 56.2 (2.6) \\
20 & 0.5 & 0.5 & 11.6 (4.1) & 8.7 (1.8)  & 72.0 (17.9) & 63.7 (6.8) \\
20 & 0.5 & 0.9 & 15.5 (9.1) & 8.6 (2.9)  & 61.4 (20.7) & 68.2 (7.5) \\
20 & 0.9 & 0.3 & 6.7 (1.8)  & 9.4 (2.3)  & 95.2 (5.6)  & 56.1 (2.3) \\
20 & 0.9 & 0.5 & 10.2 (5.6) & 8.2 (3.5)  & 64.7 (19.6) & 66.8 (7.5) \\
20 & 0.9 & 0.9 & 9.2 (2.7)  & 6.0 (3.1)  & 62.5 (20.0) & 69.3 (7.2) \\
20 & 1.0 & 0.3 & 7.9 (5.4)  & 7.2 (2.3)  & 96.7 (2.8)  & 56.0 (1.2) \\
20 & 1.0 & 0.5 & 7.1 (1.9)  & 6.6 (1.9)  & 79.3 (16.1) & 62.5 (5.7) \\
20 & 1.0 & 0.9 & 11.8 (9.2) & 5.5 (2.3)  & 69.9 (20.0) & 65.7 (7.5) \\
50 & 0.3 & 0.3 & 9.9 (3.3)  & 9.1 (2.5)  & 84.1 (12.2) & 60.0 (4.6) \\
50 & 0.3 & 0.5 & 12.1 (3.6) & 8.3 (1.0)  & 65.0 (17.3) & 67.4 (6.1) \\
50 & 0.3 & 0.9 & 12.7 (3.7) & 8.3 (1.5)  & 42.6 (20.2) & 75.2 (7.6) \\
50 & 0.5 & 0.3 &- & 8.3 (2.3)  & 88.8 (8.5)  & 58.7 (3.0) \\
50 & 0.5 & 0.5 &-& 7.1 (1.5)  & 52.8 (16.6) & 71.7 (5.7) \\
50 & 0.9 & 0.3 &-& 2.9 (0.7)  & 96.2 (4.9)  & 55.8 (1.6) \\
50 & 1.0 & 0.3 &-& 3.9 (1.4)  & 97.3 (2.9)  & 54.8 (1.4) \\
\bottomrule
\end{tabular}
\end{center}
\end{table}

\begin{table}[h]
\label{appendix:polarity}
\caption{Additional metrics for model consistency. Filtered polarity excludes updates when the agent's opinion is 0. The zero count, indicating the number of times an LLM agent selects zero, here we report both metrics in percentage of the total updates. Filtered polarity of the Zero agent is not included, as it remains constant at 0.}
\small
\begin{center}
\begin{tabular}{cccccccc}
\toprule
\multicolumn{3}{c}{} & \multicolumn{2}{c}{\textbf{\llama-2}} & \multicolumn{2}{c}{\textbf{\llama-3}} & \multicolumn{1}{c}{\textbf{Random Agents}} \\
\cmidrule(r){4-5}\cmidrule(lr){6-7}\cmidrule(l){8-8}
\(N\) & \(p\) & \(\epsilon\) & F polarity \% & Zero \%     & F polarity \%  & Zero \% & F polarity \%  \\
\midrule
20 & 0.3 & 0.3 & 2.1 (3.8)             & 11.0 (4.9)  & 1.2 (1.2)             & 10.5 (2.1)  & 46.3 (2.5)            \\
20 & 0.3 & 0.5 & 1.6 (2.0)             & 25.0 (20.4) & 1.9 (1.0)             & 27.9 (19.6) & 45.2 (2.3)            \\
20 & 0.3 & 0.9 & 2.8 (2.5)             & 50.9 (23.6) & 2.2 (2.1)             & 60.3 (23.9) & 45.8 (1.2)            \\
20 & 0.5 & 0.3 & 1.3 (1.5)             & 9.2 (4.2)   & 0.5 (0.5)             & 13.5 (6.7)  & 46.4 (2.7)            \\
20 & 0.5 & 0.5 & 2.9 (3.2)             & 31.0 (23.1) & 0.9 (1.1)             & 13.7 (13.5) & 44.9 (1.9)            \\
20 & 0.5 & 0.9 & 3.1 (4.5)             & 29.6 (22.1) & 1.2 (0.6)             & 63.9 (25.8) & 45.2 (2.3)            \\
20 & 0.9 & 0.3 & 0.6 (0.5)             & 9.8 (7.5)   & 0.2 (0.1)             & 10.8 (4.1)  & 46.3 (2.5)            \\
20 & 0.9 & 0.5 & 0.8 (0.6)             & 34.9 (26.6) & 0.6 (0.5)             & 23.9 (22.2) & 45.7 (2.0)            \\
20 & 0.9 & 0.9 & 1.0 (0.7)             & 39.3 (28.3) & 0.8 (0.5)             & 24.0 (25.9) & 45.9 (2.8)            \\
20 & 1.0 & 0.3 & 0.2 (0.3)             & 10.0 (6.1)  & 0.2 (0.2)             & 24.9 (14.9) & 45.6 (1.0)            \\
20 & 1.0 & 0.5 & 0.7 (0.7)             & 23.9 (22.0) & 0.6 (1.1)             & 14.6 (18.3) & 46.3 (1.6)            \\
20 & 1.0 & 0.9 & 1.7 (1.1)             & 27.0 (24.6) & 0.8 (0.4)             & 4.0 (1.8)   & 45.3 (1.6)            \\
50 & 0.3 & 0.3 & 0.8 (0.5)             & 20.9 (16.2) & 0.3 (0.2)             & 14.3 (10.7) & 45.2 (0.7)            \\
50 & 0.3 & 0.5 & 3.6 (3.2)             & 35.6 (22.4) & 1.2 (0.6)             & 40.0 (25.0) & 45.2 (1.4)            \\
50 & 0.3 & 0.9 & 2.4 (2.1)             & 53.5 (26.3) & 1.0 (0.5)             & 14.6 (18.4) & 45.2 (1.3)            \\
50 & 0.5 & 0.3 & - & - & 0.4 (0.4)             & 16.5 (11.7) & 45.1 (1.1)            \\
50 & 0.5 & 0.5 & - & - & 0.8 (0.3)             & 45.6 (23.1) & 45.5 (0.7)            \\
50 & 0.9 & 0.3 & - & - & 0.3 (0.4)             & 5.9 (7.4)   & 45.3 (0.6)            \\
50 & 1.0 & 0.3 & - & - & 0.2 (0.2)             & 5.3 (4.2)   & 44.6 (0.6)            \\
\bottomrule
\end{tabular}
\end{center}
\end{table}
\clearpage
\subsection{Prompt \(\theta_\text{\textbackslash n}\) results}
\label{appendix:theta_backslash}
\begin{table}[h]
\caption{Distance comparison between prompts.}
\label{appendix:theta_backslash_distance}
\small
\begin{center}
\begin{tabular}{ccccccc}
\toprule
\multicolumn{3}{c}{} & \multicolumn{2}{c}{\textbf{Prompt \(\theta_\text{base}\)}} & \multicolumn{2}{c}{\textbf{Prompt \( \theta_\text{\textbackslash n}\)}}  \\
\cmidrule(r){4-5}\cmidrule(lr){6-7}
\(N\) & \(p\) & \(\epsilon\) & \llama-2 & \llama-3     & \llama-2& \llama-3\\
\midrule
20       & 0.3 & 0.3 & 1.2 (0.3)           & 1.3 (0.2) & 0.9 (0.2)                                                & 1.3 (0.3) \\
20       & 0.3 & 0.5 & 1.2 (0.4)           & 1.1 (0.2) & 1.3 (0.3)                                                & 1.1 (0.2) \\
20       & 0.3 & 0.9 & 1.1 (0.3)           & 0.8 (0.4) & 1.1 (0.4)                                                & 0.9 (0.3) \\
20       & 0.5 & 0.3 & 1.1 (0.2)           & 1.1 (0.3) & 1.3 (0.3)                                                & 1.2 (0.3) \\
20       & 0.5 & 0.5 & 1.0 (0.4)           & 1.1 (0.4) & 1.2 (0.4)                                                & 1.1 (0.3) \\
20       & 0.5 & 0.9 & 1.4 (0.4)           & 0.8 (0.4) & 1.4 (0.4)                                                & 1.2 (0.4) \\
20       & 0.9 & 0.3 & 0.9 (0.2)           & 1.1 (0.2) & 0.8 (0.2)                                                & 1.2 (0.2) \\
20       & 0.9 & 0.5 & 1.0 (0.3)           & 1.0 (0.4) & 1.0 (0.4)                                                & 1.3 (0.2) \\
20       & 0.9 & 0.9 & 1.0 (0.3)           & 1.6 (0.5) & 1.4 (0.5)                                                & 1.3 (0.5) \\
20       & 1.0 & 0.3 & 1.0 (0.2)           & 1.0 (0.3) & 1.3 (0.3)                                                & 1.4 (0.3) \\
20       & 1.0 & 0.5 & 1.0 (0.4)           & 0.8 (0.4) & 0.9 (0.3)                                                & 0.9 (0.2) \\
20       & 1.0 & 0.9 & 1.2 (0.4)           & 2.1 (0.4) & 1.1 (0.4)                                                & 1.2 (0.4) \\ 
\bottomrule
\end{tabular}
\end{center}
\end{table}

\begin{table}[h]
\label{appendix:prompt_backslash_polarity}
\caption{Polarity percentage comparison between prompts.}
\small
\begin{center}
\begin{tabular}{ccccccc}
\toprule
\multicolumn{3}{c}{} & \multicolumn{2}{c}{\textbf{Prompt \(\theta_\text{base}\)}} & \multicolumn{2}{c}{\textbf{Prompt \( \theta_\text{\textbackslash n}\)}}  \\
\cmidrule(r){4-5}\cmidrule(lr){6-7}
\(N\) & \(p\) & \(\epsilon\) & \llama-2 & \llama-3     & \llama-2& \llama-3\\
\midrule
20 & 0.3 & 0.3 & 14.4 (8.3)          & 11.8 (2.0) & 13.3 (8.4) & 16.7 (2.7) \\
20 & 0.3 & 0.5 & 13.5 (5.9)          & 14.5 (3.0) & 20.0 (8.9) & 17.2 (4.0) \\
20 & 0.3 & 0.9 & 15.7 (7.0)          & 10.8 (2.2) & 22.2 (9.8) & 16.1 (4.4) \\
20 & 0.5 & 0.3 & 9.2 (3.7)           & 11.2 (3.9) & 9.3 (5.6)  & 13.8 (1.8) \\
20 & 0.5 & 0.5 & 11.6 (4.1)          & 8.7 (1.8)  & 10.9 (5.4) & 11.4 (1.7) \\
20 & 0.5 & 0.9 & 15.5 (9.1)          & 8.6 (2.9)  & 13.0 (6.9) & 12.3 (3.5) \\
20 & 0.9 & 0.3 & 6.7 (1.8)           & 9.4 (2.3)  & 3.7 (1.8)  & 9.6 (2.8)  \\
20 & 0.9 & 0.5 & 10.2 (5.6)          & 8.2 (3.5)  & 9.4 (3.6)  & 11.2 (3.4) \\
20 & 0.9 & 0.9 & 9.2 (2.7)           & 6.0 (3.1)  & 7.5 (3.0)  & 8.4 (3.5)  \\
20 & 1.0 & 0.3 & 7.9 (5.4)           & 7.2 (2.3)  & 6.0 (2.3)  & 11.1 (3.6) \\
20 & 1.0 & 0.5 & 7.1 (1.9)           & 6.6 (1.9)  & 6.6 (3.0)  & 7.8 (1.9)  \\
20 & 1.0 & 0.9 & 11.8 (9.2)          & 5.5 (2.3)  & 11.7 (6.9) & 5.5 (2.0)  \\

\bottomrule
\end{tabular}
\end{center}
\end{table}
\end{document}